\documentclass[12pt]{article}
\usepackage{graphicx}
\begin{document}
\begin{center}
{\LARGE \bf Stellar Structure of Irregular Galaxies. I. Face-on Galaxies}
\bigskip

{\large \bf N. A. Tikhonov}

\bigskip

{\em Special Astrophysical Observatory, Russian Academy of Sciences, Nizhnii Arkhyz, Karachai-Cherkessian Republic, 357147 Russia}
\end{center}

\bigskip

Received January
29, 2004; in final form, February 17, 2005
\bigskip

{\large \bf Abstract} \\

\bigskip

Stellar photometry of nearby irregular galaxies of the Local Group is used to
identify and study the young
and old stellar populations of these galaxies. An analysis of the spatial
distributions of stars of different ages in face-on galaxies shows that the
young stellar populations in irregular galaxies are concentrated toward the
center, and form local inhomogeneities in star-forming regions, while the old
stellar populations--red giants--form extended structures around the irregular
galaxies. The sizes of these structures exceed the visible sizes of the galaxies
at the 25$^m$/arcsec$^2$ isophote by a factor of two to three. The surface density of
the red giants decreases exponentially from the center toward the edge, similar
to the disk components in spiral galaxies.

\section{INTRODUCTION}
Dwarf irregular galaxies are the most numerous of all
observed galaxies. Their absolute luminosities lie in the interval from
$M_B = -9^m$ to $M_B = -18^m$. Such galaxies usually have higher hydrogen
contents than other types of dwarf galaxies, and are host to sometimes very
intense star-forming processes. A very energetic burst of star formation may
promote a galaxy to the class of blue compact dwarf galaxies. Many bright
irregular galaxies exhibit signs of weak spiral arms, which are especially
prominent in the distribution of neutral hydrogen, as is the case, e.g.,
in HoII [1] and IC 10 [2].

In this paper, we do not study normal spiral
galaxies, which have higher masses and rotational velocities than irregular
galaxies, although the dividing line between irregular and spiral galaxies
is fairly arbitrary.

In his classification [3], Hubble introduced the
class of irregular galaxies as a sort of a "wastebasket" for all the
objects that could not be classed as spirals, ellipticals, or lenticulars.
As is clear from their name, irregular galaxies have no regular shape.
However, much observational data has been collected since the Hubble
classification of galaxies was first adopted, and many of the initial
views about the morphology and physical parameters of galaxies have
evolved.

Hodge and Hitchcock [4] were the first to suggest, based on
measurements of the apparent sizes of 131 galaxies, that irregulars may
have flattened shapes. Some authors had expressed similar ideas before
[3], but produced no corroborative observational evidence. Later, radio
observations established that irregular galaxies rotate, thereby
putting them together with spiral galaxies in the group of axisymmetric
objects. However, the spatial stellar structure of irregular galaxies
remained unclear, since their images exhibited chaotically scattered
regions of bright young supergiants, while the old stellar populations
were detectable only near the telescope limits in Local-Group galaxies.

A new era in the study of galaxies began in the 1980s with the advent
of new detectors with high quantum efficiency--CCDs. Surface photometry
has established that irregular galaxies show exponential radial brightness
decreases [5, 6] similar to those observed for the disk components in
spiral galaxies, suggesting a similarity between spiral and irregular
galaxies.

Photoelectric color measurements of galaxies of various types
show that irregular galaxies become redder toward their edges. This was
easy to understand, since it was already known even at the time of Baade
[7] that the peripheral regions of irregular galaxies in the Local Group
are populated by a considerable number of red giants, which are visible
on the Palomar plates. This suggested that other irregular galaxies might
have a similar structure. The distribution of red giants in irregular
galaxies is often described using the term "sheet," introduced by Baade,
or "halo," used in modern papers. The use of different terms leads to
confusion in descriptions of galaxy morphology. Whereas the term "sheet"
means simply a stellar distribution with a low surface-density gradient,
"halo" implies a spherical distribution of stars around the associated
galaxies.

We believe that the use of the term "halo" for irregular
galaxies is incorrect, because, as we show below, most irregular galaxies
possess only an extended thick disk, which can be mistaken for a halo.
Only massive galaxies have a genuine more extended halo in addition to
the thick disk.

\begin{table}
\caption{Face-on galaxies}
\bigskip
\begin{tabular}{|rllclc|} \hline
    &                &                       &     &          &         \\
 N  &   Name          &   R.A. (2000.0) Dec.    & Type & Size,(arcmin)   & Magnitude \\
    &                &                       &     &          &         \\
\hline
    &                &                       &     &          &         \\
 1  &  K52           &  08 23 56.0 $+$71 01 46 & I ? &1.3$\times$  0.7 &  16.5
 \\
 2  &  DDO53         &  08 34 07.2 $+$66 10 54 & Im  &1.5$\times$ 1.3 &  14.5
\\
 3  &  K73           &  10 52 55.1 $+$69 32 47 & Im  &0.6$\times$ 0.4 &  14.9
\\
 4  &  GR8           &  12 58 40.4 $+$14 13 03 & ImV &1.1$\times$ 1.0 &  14.7
\\
 5  &  HIPASS1321-31 &  13 21 08.2 $-$31 31 45 & Im  &0.7$\times$ 0.5 &  17.0
\\
 6  &  DEEP1337-33   &  13 37 04.3 $-$33 21 51 & Im  &0.3$\times$ 0.3 &  16.8
\\
 7  &  PGC48111      &  13 37 20.0 $-$28 02 42 & Im  &1.3$\times$ 1.0 &  15.0
\\
 8  &  HIPASS1337-39 &  13 37 26.0 $-$39 53 47 & Im  &0.6$\times$ 0.5 &  16.1
\\
 9  &  UGC8833       &  13 54 48.7 $+$35 50 15 & Im  &0.9$\times$ 0.8 &  16.5
\\
 10 &  DDO187        &  14 15 56.5 $+$23 03 19 & ImV &1.7$\times$ 1.3 &  14.4
\\
 11 &  DDO190        &  14 24 43.4 $+$44 31 33 & IAm &1.8$\times$ 1.8 &  13.2
\\
 12 &  UGCA438       &  23 26 27.5 $-$32 23 20 & IBm &1.5$\times$ 1.2 &  13.9
 \\
    &                &                       &     &          &         \\
\hline
\end{tabular}
\end{table}

Deep images have been taken for all galaxies in the Local Group and a few dozen
other nearby galaxies. These show both bright, young supergiants and fainter red
 giants. The spatial distribution of stars of various types has been studied in
several irregular galaxies [8--14].

The most detailed of these studies showed
that:

(a) The distribution of red giants extends well beyond the 25$^m$/arcsec$^2$ limit
commonly used to define the galaxy sizes;

(b) In the few cases when the
corresponding measurements were made, the density of red giants was found
to decrease exponentially from the center of the galaxy towards the edge,
similar to the disk components in spiral galaxies.

Minniti and Zijlstra [8, 12]
interpreted the extended distributions of red giants in the galaxies WLM and
NGC 3109 as evidence for halos, i.e., spheroidal distributions of stars. However,
this conclusion is not justified, since they did not report the measured
distribution of red giants along the major axis of the galaxy. Establishing
the real spatial distribution of stars requires detailed studies of individual
galaxies, and analyses of a representative sample of galaxies are needed to
eliminate effects due to features associated with individual galaxies.

\section{SELECTION OF OBJECTS AND OBSERVATIONS}
 Studies of the spatial structures of flat systems require observations of both
edge-on and face-on galaxies of the same type. The shapes of edge-on
galaxies in the $Z$ direction can easily be determined, and the disk and halo, if
any, identified. Moreover, in such galaxies the boundaries of the distributions
of stars of various types are easier to determine, since they are projected onto
a small area of sky.

Observations of face-on galaxies can be used to analyze the
distribution of stars from the center to the periphery, which is not possible
for edge-on galaxies. Thus, joint studies of galaxies viewed at different angles
make it possible to study the spatial distributions of the constituent stars.
Other criteria used to select our target objects included their rotational
velocities ($V_r < 100$ km s$^{-1}$), morphological types (Irr, Irr/Sph), distances
($D < 5$ Mpc), and sizes. We used the rotational velocities as a check for
their morphological types; i.e., irregular galaxies are known to have relatively
low rotational velocities. We compiled a list of galaxies meeting our criteria
and cross-identified it with the HST archive. Resolving galaxies beyond the Local
Group into stars is difficult for ground-based telescopes, except for the
brightest supergiants, whereas this is easy with the high spatial resolution
of the HST.

In addition to the images adopted from the archive of the Hubble
Space Telescope, we used images of the three irregular Local-Group galaxies
IC 10, IC 1613, and Pegasus obtained by us in 1999 with the 6-m telescope of
the Special Astrophysical Observatory (SAO). In addition, the galaxies DDO 187,
 DDO 190, DDO 210, and SagDIG were observed in 1997 with the 2.5-m Nordic
Optical Telescope (NOT).

\begin{figure}[h]
\centerline{
\includegraphics[width=8.cm,angle=0,clip]{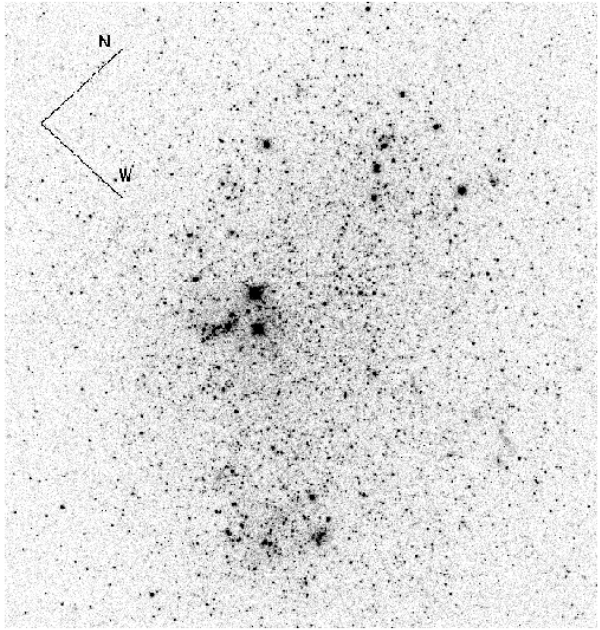} }
\caption{Central part (72$^{\prime\prime}\times77^{\prime\prime}$ ) of GR8 as seen in the F550w and F814w
HST images. The young, bright stars are concentrated in star-forming
regions, whereas the fainter stars (mostly red giants) show a more extended
and smoother structure.}
\end{figure}

Observations with the SAO 6-m telescope were made
with a CCD with a 2.52$^{\prime}\times2.5^{\prime}$ field of
view for the image scale of 0.137$^{\prime\prime}$/pixel. Observations
at the NOT were made with the HiRAC camera
equipped with a Loral CCD producing a 3.75$^{\prime}\times3.75^{\prime}$
field of view for the
image scale of 0.11$^{\prime\prime}$/pixel. The HST field is
2.5$^{\prime}$ in size and has an image
scale of 0.1$^{\prime\prime}$/pixel. For many galaxies, these fields of view are too small
to enable photometry of stars in the entire galaxy, but the required data
about the stellar structure can be derived by selecting small galaxies or
galaxies with efficiently chosen fields at appropriate locations along the
major and minor axes. Our final list included 12 face-on and 12 edge-on
galaxies. In addition to the irregular galaxies, we also studied about
twenty galaxies of other types, however, those results lie beyond the scope
of this paper.

Because of the large amount of observational data analyzed,
we restrict our analysis here to the face-on galaxies. The table lists these
galaxies and their main parameters, adopted from the NED database. The results
for the two additional galaxies IC 10 and IC 1613 can be found in [15, 16].

\section{STELLAR PHOTOMETRY}
 Prior to carrying out photometry of individual stars, we performed a
standard preliminary reduction
of all the images, including dark-current subtraction, elimination of
irregularities of the CCD sensitivity by dividing the image by a flat
field taken at dusk, and removal of cosmic-ray traces and "hot" pixels.

\begin{figure}[h]
\centerline{
\includegraphics[width=9.cm,angle=-90,clip]{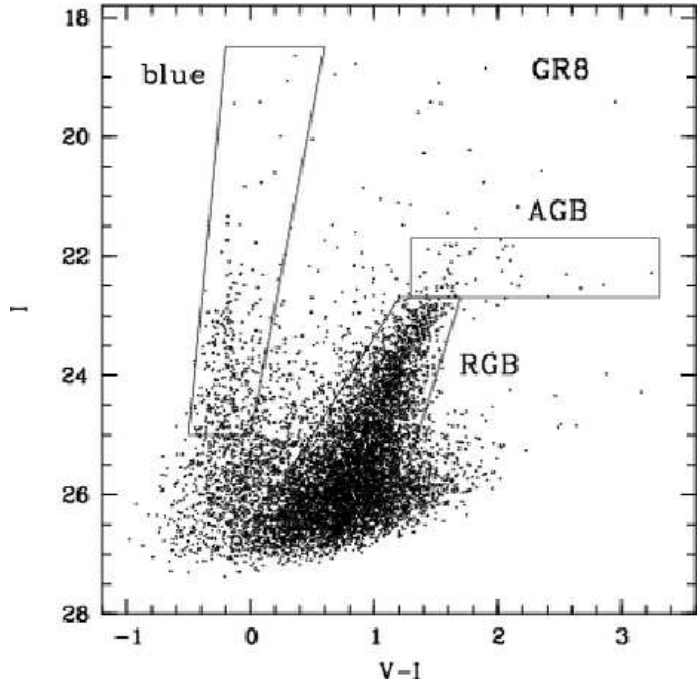} }
\caption{Color--Magnitude diagram for stars of the galaxy GR8. Regions of young
("blue"--blue supergiants and giants), intermediate (``AGB''--asymptotic giant
branch), and old ("RGB"--red giant branch) stars are marked.}
\end{figure}

\begin{figure}[h]
\centerline{
\includegraphics[width=9.cm,angle=-90,clip]{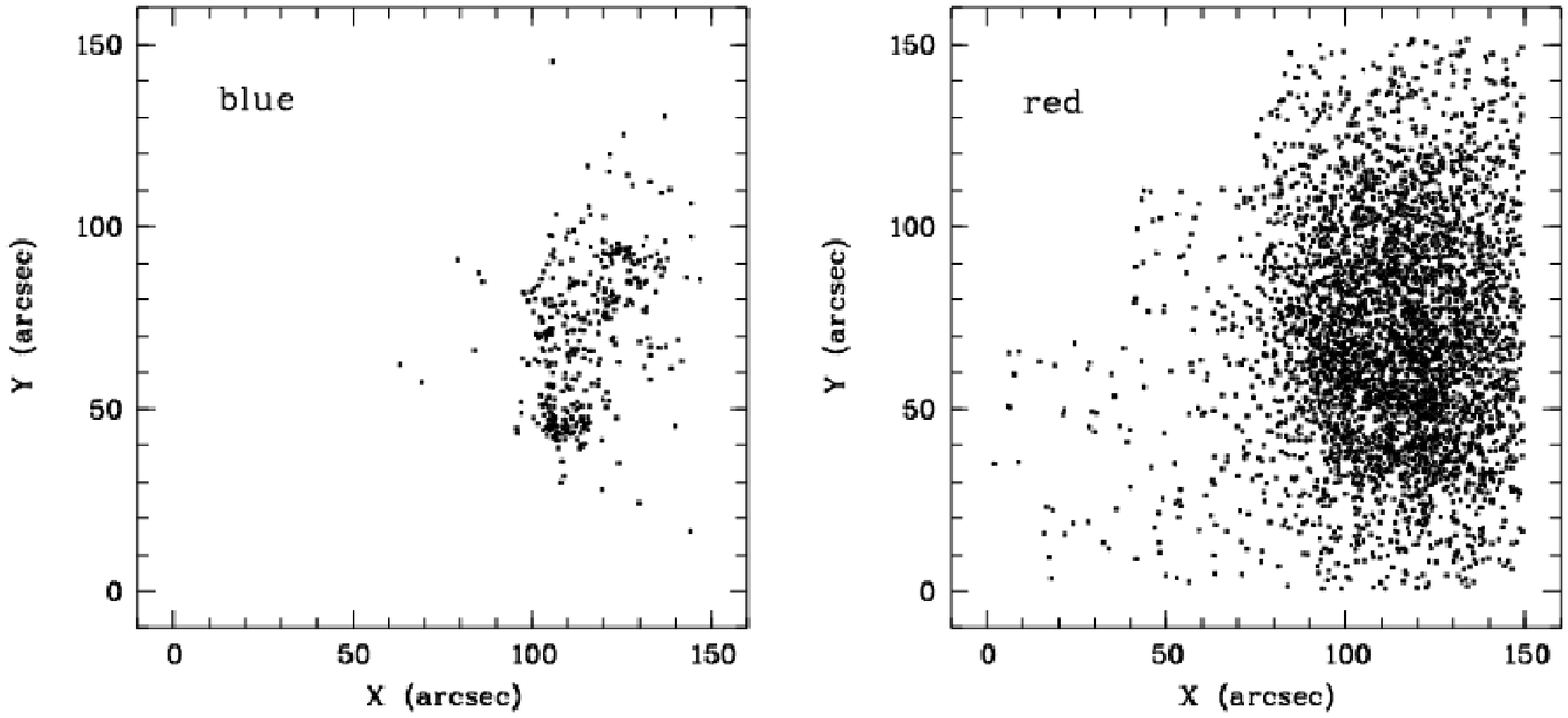} }
\caption{Distributions of young (blue supergiants) and old (red giants)
 stars in GR8. All the young stars are concentrated near the central
region of the galaxy, whereas the red giants form an extended structure
resembling a thick disk.}
\end{figure}

The images obtained from the HST archive did not require this procedure,
which was performed automatically when we ordered the images from the archive.
We extracted data for the stars and carried out the photometric measurements
using the standard DAOPHOT II procedure in the MIDAS package [17]. To translate
the instrumental magnitudes into the Kron--Cousins system, we carried out
photometric observations of standard stars from the list of Landolt [18] on
each night to derive the equations for the translation from the instrumental
magnitudes to the $BVRI$ system.

The photometric measurements produced a list
of coordinates, magnitudes, and accompanying parameters for the stars, which
could be used to assess the photometric accuracy and differences of the
photometric profiles of the objects studied from a standard stellar profile.

\section{MEASUREMENT OF THE SURFACE DENSITY OF STARS}
Figure 1 shows GR8, which is a typical irregular galaxy in our list.
Figure 2 shows the Hertsprung--Russell diagram for this galaxy based
on our stellar photometry, with regions occupied by different stellar
types marked. We constructed distributions of various types of stars
projected onto the plane of the galaxy (Fig. 3) and calculated the surface
density of stars within annuli centered on the galaxy. We made the
necessary corrections for the incompleteness of the stellar sample due to
the asymmetric distribution of the four chips of the HST field relative to
the center of the galaxy.

Figure 4 shows the results for the 12 galaxies
in the form of plots of the density distributions for young (blue and red
supergiants) and old (red giants) stars.

Star-forming processes that produce
bright supergiants occur in all irregular galaxies. The presence of bright
stars and the nonuniform background brightness distribution in star-forming
regions hinder photometry of relatively faint stars, including red giants.
As a result, the density of red giants is underestimated in the central
regions of galaxies, as can seen in Fig. 4. Another effect that decreases
the calculated stellar density in central regions of galaxies is the
"blending" of stellar images due to the extreme crowdedness of the stars
in the central regions of galaxies. The DAOPHOT program for stellar photometry
removes such images from the list, interpreting them as being nonstellar.
This decreases both the completeness of the sample and the calculated
surface density of the stars, especially in the central regions of
galaxies. We were able to compensate for these effects using the
standard procedure of artificial-star photometry. However, the statistical
influence of these effects is insignificant for stars at the periphery,
and we excluded the parameters of the stellar distributions near the
galactic centers when deriving the parameters for the exponential radial
decrease of the stellar density.

\begin{figure}[h]
\centerline{
\includegraphics[width=14.cm,angle=0,clip]{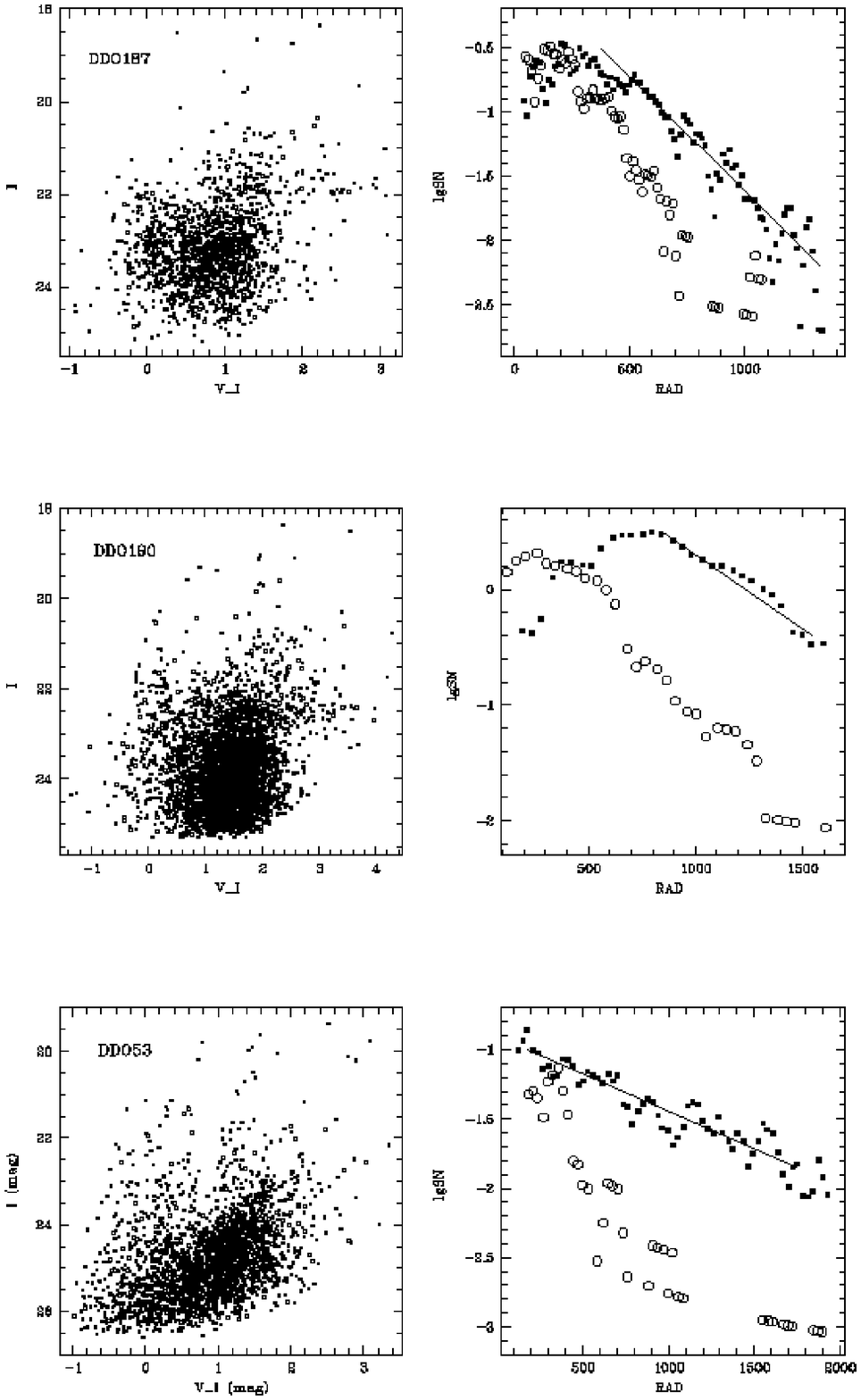} }
\caption{Color--magnitude diagrams and radial distributions of stellar density
in irregular galaxies. All the color--magnitude diagrams show both young, blue
stars (open circles) and stars with ages of several billion years (filled squares).
On the radial distribution plots, young stars occupy the visible regions of the galaxies,
whereas the red giants, whose density decreases exponentially from
the center, can easily be traced to much greater galactocentric distances.}
\end{figure}

\section{STRUCTURE OF THE IRREGULAR GALAXIES}
The results shown in Fig. 4 demonstrate that, despite the seemingly
chaotic nature of the distribution of star-forming regions, they are
nonetheless appreciably concentrated toward the centers of their galaxies,
as can be seen in the density distributions for the young stars. It is
obvious from a comparison of the morphologies of galaxies with different
masses that  the star-forming regions
in more massive galaxies begin to be concentrated into spiral-arm type
structures, which, however, remain barely distinguishable.

While the young,
blue stars are concentrated in star-forming regions, giving the galaxy its
irregular appearance, the red giants have a smoother distribution. Their
surface-density contours accurately indicate the center of the galaxy, and
can be used to estimate its inclination to the line of sight.
\setcounter{figure}{3}
\begin{figure}[h]
\centerline{
\includegraphics[width=14.cm,angle=0,clip]{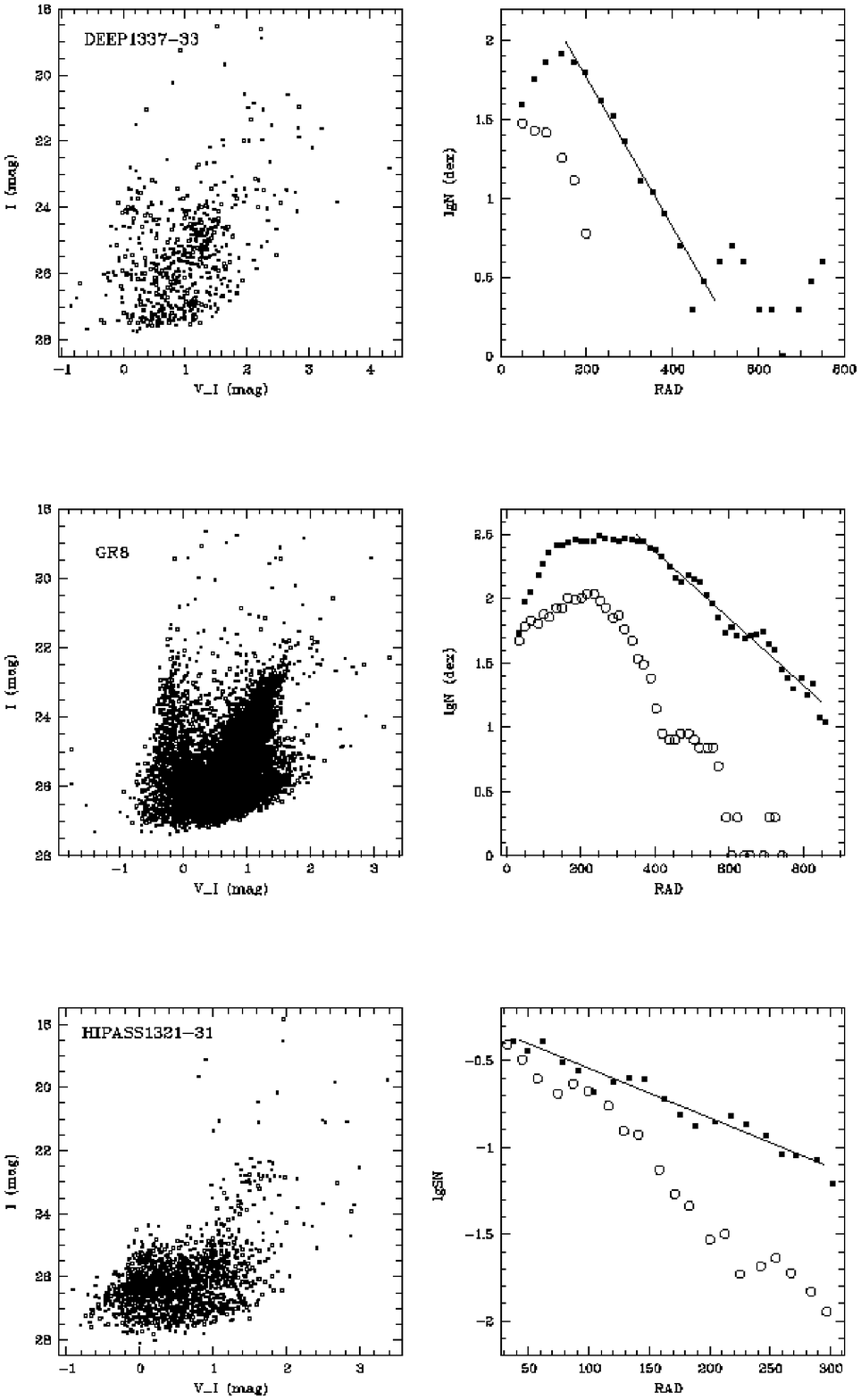} }
\caption{(Contd.)}
\end{figure}
In all the galaxies, the red supergiants extend much farther from the center
than the blue giants (Fig. 4). The ratio of the sizes of the corresponding spatial
structures can reach five (IC 10), but its mean value is 2.5--3.0. The domains
occupied by red giants have fairly welldefined boundaries, with very few, mostly
foreground, stars seen beyond them. An analysis of the density distribution of
red giants in IC 10 revealed a thick disk with a sharp boundary [15]. An even
larger structure extending beyond the thick disk has a different density
gradient than the thick disk. Thus, this galaxy
exhibits a complex structure consisting of a thick disk and halo, similar
to the structure of spiral galaxies [19, 20]. IC 10 differs from the other
irregular galaxies in our list only in its luminosity (mass), suggesting that
it is a galaxy's mass that determines whether it has a genuine halo. Although
the small numbers of such galaxies prevent us from drawing firm conclusions,
this observational fact can, nevertheless, be easily explained if there is
an overall population of disk galaxies, whose low-mass and high-mass members
have come to be known as irregulars and spirals,
respectively.
\setcounter{figure}{3}
\begin{figure}[h]
\centerline{
\includegraphics[width=14.cm,angle=0,clip]{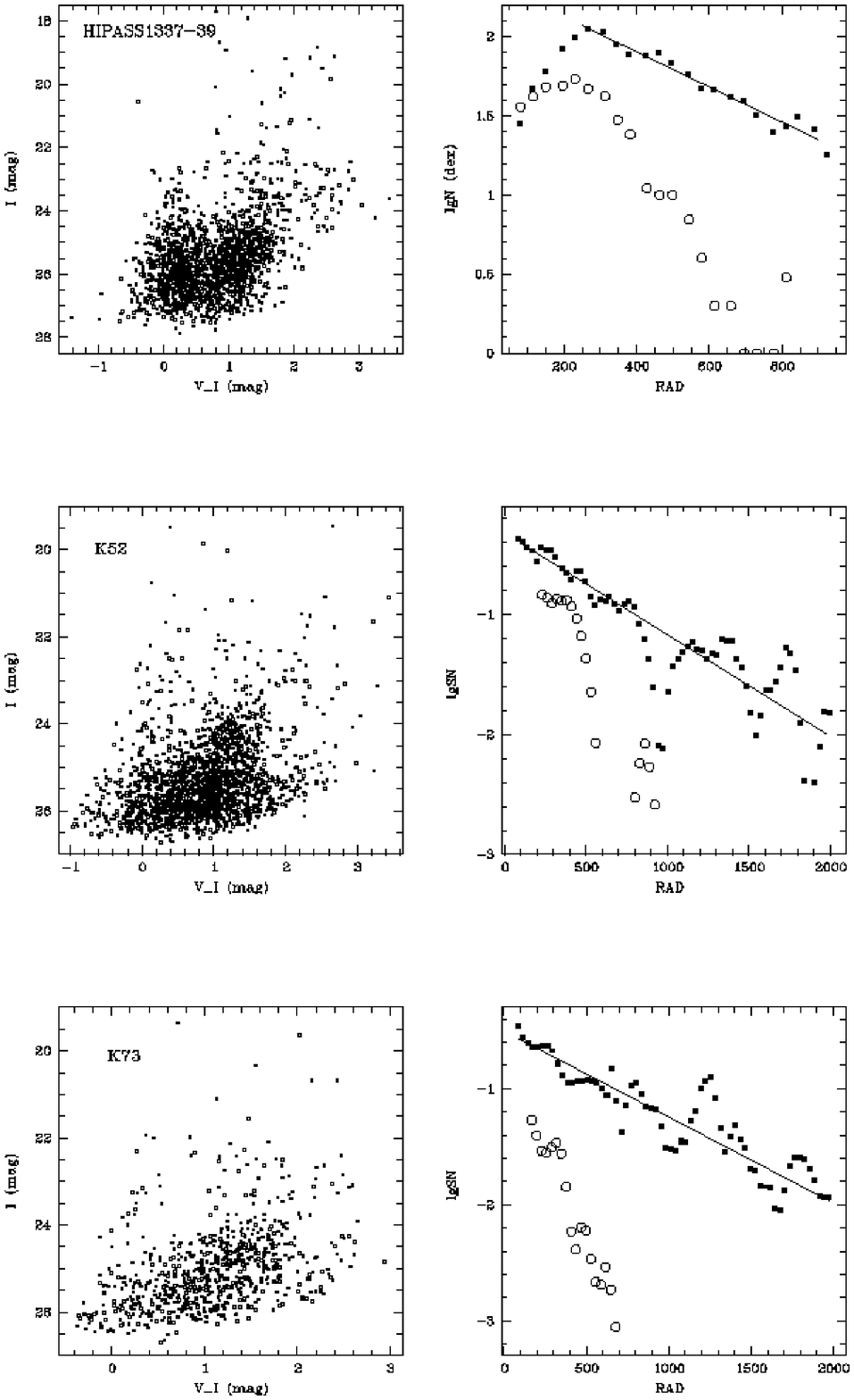} }
\caption{(Contd.)}
\end{figure}
The reality of such a unified scheme is supported by the continuous range
of physical parameters for irregular and spiral galaxies. In this case, the
presence of a halo around a galaxy is determined only by the mass of the galaxy,
and not by its membership in some artificially defined class of irregular or
spiral galaxy.

The stars at the peripheries of galaxies have stable orbits,
implying that all evolved invisible stars should be located within the boundaries
outlined by the current distribution of red giants. In other words, if the dark
 matter in galaxies consists of the remnants of stellar evolution, the boundaries
of the distribution of this matter can be determined by analyzing the distribution
of red giants.

\section{CONCLUSIONS}
Our analysis of the stellar populations in 12
irregular face-on galaxies has led to the following conclusions.

(1) The radial distributions of stars of different types in the galaxy
show exponential decreases, similar to the behavior of a disk component.

\setcounter{figure}{3}
\begin{figure}[h]
\centerline{
\includegraphics[width=14.cm,angle=0,clip]{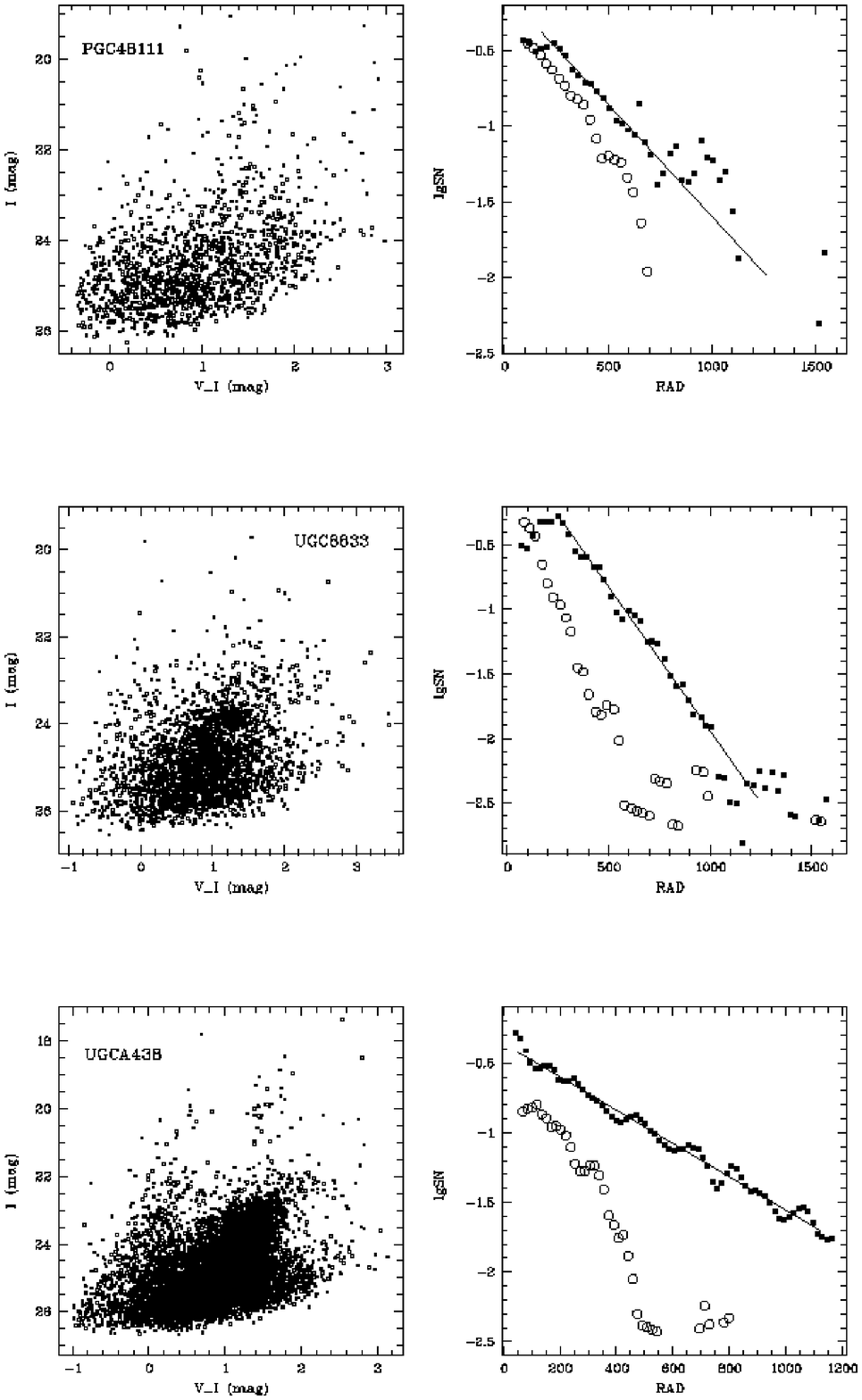} }
\caption{(Contd.)}
\end{figure}

The rate at
which the stellar density in a given galaxy decreases depends on the age of
the stars. Younger stars show more rapid density decreases toward the periphery
than older stars.

(2) The stellar populations of galaxies have fairly sharp boundaries if these
 boundaries coincide with the cutoff of the disk formed by red giants.

(3) Our stellar density measurements show that old stars form disks that extend
a factor of two to three further than the visible isophotal sizes at 25$^m$/arcsec$^2$.

(4) All these conclusions refer to dwarf galaxies. Our preliminary studies show
that massive irregular galaxies may possess a more extended genuine halo, in
addition to a thick disk, as is observed in IC 10.

\bigskip

{\bf ACKNOWLEDGMENTS}
This work was supported by the Russian Foundation for Basic Research
(project nos. 00--02--16584 and 03--02--16344). Our work has made use of data
from the NASA/IPAC Extragalactic Database.

\bigskip

{\bf REFERENCES}

1. D. Pushe, D. Westphahl, et al., Astron. J. 103, 1841 (1992).

2. E. Wilcots and B. Miller, Astron. J. 116, 2363 (1998).

3. A. Sandage, The Hubble Atlas of Galaxies (Carnegie Inst., Washington, 1961).

4. P. Hodge and J. Hitchcock, Publ. Astron. Soc. Pac. 78, 79 (1966).

5. T. Bremens, B. Bingelli, and P. Prugniel, Astron. Astrophys., Suppl. Ser. 129, 313 (1998).

6. T. Bremens, B. Bingelli, and P. Prugniel, Astron. Astrophys., Suppl. Ser. 137, 337 (1998).

7. W. Baade, Evolution of Stars and Galaxies (Harvard Univ. Press, Cambridge, 1963).

8. D. Minniti and A. Zijlstra, Astron. J. 114, 147 (1997).

9. A. Aparicio, N. Tikhonov, and I. Karachentsev, Astron. J. 119, 177 (2000).

10. R. Lynds, E. Tolstoy, E. J. O'Neil, and D. Hunter, Astron. J. 116, 146 (1998).

11. D. Martinez-Delgado, K. Gallart, and A. Aparicio, Astron. J. 118, 862 (1999).

12. D. Minniti, A. Zijlstra, and V. Alonso, Astron. J. 117, 881 (1999).

13. A. Aparicio and N. Tikhonov, Astron. J. 119, 2183 (2000).

14. B. Letarte, S. Demers, P. Battinelli, and W. Kunkel, Astron. J. 123, 832 (2002).

15. N. A. Tikhonov, Doctoral Dissertation (St. Petersburg, 2002).

16. N. Tikhonov and O. Galazutdinova, Astron. Astrophys. 394, 33 (2002).

17. P. Stetson, Users Manual for DAOPHOTII (1994).

18. A. U. Landolt, Astron. J. 104, 340 (1992).

19. J. Guillandre, J. Lequeux, and L. Loinard, in IAU Symposium No. 192:
The Stellar Content of Local Group Galaxies, Ed. by P. Whitelock and R. Cannon
(1988), p. 27.

20. N. Tikhonov and O. Galazutdinova, Astron. Astrophys. (2004) (in press). Translated by A. Dambis

\end{document}